\documentclass[letterpaper,10pt]{article}
\usepackage[T1]{fontenc}
\usepackage{newtxtext,newtxmath}
\usepackage[hyphens]{url}
\usepackage{graphicx}
\usepackage{amsmath}
\usepackage{array}
\usepackage{booktabs}
\usepackage{caption}
\usepackage{placeins}
\newcommand{\PaperTitle}{High-Coverage Blind Spots in Online Abuse Monitoring: Account Selection, Held-Out Flagging Rates, and Sample Composition}
\title{\bfseries\PaperTitle}
\author{Ziming Liu \\ The University of Melbourne}
\date{}
\begin{document}
\twocolumn[
\begin{@twocolumnfalse}
\begin{center}
{\LARGE\bfseries\PaperTitle\par}
\vspace{.1in}
{\large\bfseries Ziming Liu}\par
{\normalsize The University of Melbourne}\par
\end{center}
\vspace{.10in}
\begin{abstract}

When monitoring online abuse, platforms may select accounts for observation based either on the absolute number of abusive replies received or on the proportion of abusive replies relative to the total volume of replies. These two selection methods can yield very different sets of accounts. High-profile accounts, which receive a large volume of replies, naturally accumulate a higher total count of abusive responses. Conversely, an ordinary account---even with a small total number of replies---may exhibit a very high proportion of content flagged as abusive. We investigate what is overlooked when selecting accounts based on absolute counts, and whether accounts selected based on proportions maintain a high labeling rate when examining their broader reply history.

We analyzed data on Online Abusive Attacks (OAA), comprising 945,987 replies across 384 accounts, of which 19,152 were flagged as abusive. When selecting a fixed set of 25 accounts, the lists generated by absolute count versus smoothed labeling rate share only one account in common, covering 75.56\% and 5.04\% of the total flagged replies, respectively. In a null-model simulation (3,000 iterations) that preserved actual reply volumes and eight account groups, the overlap remained minimal---averaging only 0.5 accounts---while the ratio of labeling rates (the "rate multiplier") averaged 0.916. In a 500-iteration stratified hold-out comparison using the actual dataset, the labeling rates for the two lists were 2.69\% and 4.98\%, respectively. Across 3,000 resamplings of the account set, the 95\% percentile interval for this rate multiplier was 1.38--2.18.

While the two selection methods produce distinct lists in the null model, the "proportion-based" list does not exhibit a higher flagging rate in that context; however, in the actual dataset, the proportion-based list retains a higher held-out flagging rate. The proportions of the "Identity\_Attack" category in the two lists are 1.94\% and 18.01\%, respectively. Consequently, selecting accounts based on the total number of flagged replies versus the proportion of flagged content yields not only different lists of accounts but different distributions of flagged-reply categories. An independent UK election dataset also yields divergent account lists, while rate differences vary with group composition and evaluation period.


\end{abstract}
\vspace{.08in}
\end{@twocolumnfalse}
]

\section{Introduction}

When monitoring for online attacks, platforms must first decide which accounts to prioritize for their limited monitoring capacity. One approach is to select accounts based on the total number of abusive replies they receive, while another is to select them based on the proportion of abusive replies relative to the total volume of replies. Although both methods utilize the same dataset, the resulting lists of accounts can differ significantly.

Consider Account A, which receives 10,000 replies, 200 of which are flagged as abusive---a proportion of 2\%. Now consider Account B, which receives 500 replies, 50 of which are flagged as abusive---a proportion of 10\%. If selection is based on the absolute number of flagged replies, Account A would rank higher; if based on the proportion, Account B would rank higher.

An account that receives a higher volume of replies is more likely to accumulate a larger total number of abusive replies. For instance, a high-profile account might receive a substantial number of abusive replies, even if the proportion of such replies remains low. Conversely, a typical account might receive only 100 replies, yet if 50 of them are flagged as abusive, the proportion reaches 50\%. Therefore, if the goal is to determine the extent to which an account's received replies are abusive, comparing them by proportion is more direct than simply looking at the total count of abusive replies.

By selecting accounts for monitoring based on the total number of abusive replies, a platform can identify a larger volume of existing abusive content. However, this approach may overlook "ordinary" accounts that receive fewer total replies but have a high proportion of aggressive ones. For such accounts, even if the absolute number of abusive replies is small, it may still indicate that a significant portion of the replies they receive are abusive in nature. We seek to understand the severity of this omission and whether the higher flagging rates observed for certain accounts persist when analyzed against a different subset of historical replies.

Existing research provides context for these questions. Alharthi et al. examined the distribution of abusive replies across accounts and original posts {[}1{]}; Burke-Moore et al. compared the "abuse burden" faced by different groups of public figures and its relationship to account characteristics {[}2{]}; and Bandy and Lazovich distinguished between the absolute volume and the proportion of abusive content encountered by users, finding that both metrics vary depending on how content is ranked {[}3{]}. While these studies address the distribution of aggression and the measurement of exposure, our research focuses on a different question: when observation resources are constrained by a fixed number of account slots, which accounts are selected for monitoring, and are there repeatable differences in held-out flagging rates between selected accounts?

Differences in the resulting lists do not automatically imply that accounts selected based on their flagging *proportion* will consistently exhibit higher flagging rates. For instance, an account might show a 10\% flagging rate in one part of its historical replies; when evaluated against another part of those replies, that rate could remain 10\% or drop significantly. To test this, we randomly split each account's total count of flagged and unflagged replies into two subsets: one used for selection and the other for evaluating flagging rates. We also developed a null model---incorporating the actual volume of replies and eight account groups---that assumes accounts within the same group share an identical probability of being flagged. This allows us to determine whether the observed differences in account selection could be explained solely by reply volume and random variation. In our analysis of 384 accounts (with a fixed capacity of 25 slots), only one account appeared in both lists generated from the full dataset. However, in our hold-out validation using historical OAA data, the accounts selected based on their flagging proportion continued to exhibit a flagging rate approximately 1.85 times higher than the others.

\section{Related Work}

\subsection{Aggression Burden, Opportunity for Interaction, and Targets of Aggression}

The Online Abusive Attacks (OAA) dataset links accounts, original posts, and the replies they receive, enabling the aggregation of abusive content by the target of the replies {[}1{]}. Burke-Moore et al. compared the aggression experienced by UK Members of Parliament, professional footballers, and journalists, highlighting the differences between absolute counts of aggression and levels adjusted for relevant confounding factors; they also found that aggression is highly concentrated among a small number of public figures {[}2{]}. These studies demonstrate that an account's opportunity to be discussed and the "abuse burden" it faces cannot be treated as the same metric. In our fixed-capacity monitoring scenario, this distinction further influences which accounts are ultimately included in the observation sample.

\subsection{From content prevalence to monitoring selection with limited slots}

Bandy and Lazovich examined users' actual exposure to marginally abusive content within their timelines. They found that even when the proportion of such content is relatively low, cumulative exposure can still increase as the total volume of content consumed grows {[}3{]}. This demonstrates that the absolute quantity of content and its relative proportion can move in opposite directions. While their study focused on individual users exposed to content, the present study focuses on the accounts that receive replies. By imposing a fixed quota on the number of accounts to be monitored, the two measurement approaches---count-based versus proportion-based---necessarily result in different sets of target accounts.

Park, Seering, and Bernstein employed manual annotation, model estimation, and sampling to study the incidence of policy-violating comments within online communities {[}4{]}. Milne, McCabe, and Calvo utilized automated triage in online support communities and observed subsequent changes in moderator responses {[}5{]}. Estimating the incidence of problematic content and determining moderator responses are distinct tasks; between them lies a critical, yet often overlooked, step: deciding which specific entities should receive attention given limited resources. In this paper, we use an intra-group homogeneity null model and a hold-out dataset to test whether the differences in the lists generated by this allocation process correspond to repeatable differences in held-out flagging rates.

\section{Data and Comparative Design}

\subsection{Data and Unit of Analysis}

This study utilizes the account-level summary CSV file made public by the original authors of the Online Abusive Attacks (OAA) dataset {[}1,6{]}, corresponding to Git commit version 75ee52a947c94687471be4ad46f914157192fed4. The source file contains 2,368 rows, with some accounts appearing multiple times. We retained only the first valid record for each account, resulting in 1,780 unique accounts; of these, 1,088 received at least one reply. For our primary experiment, we selected the 384 accounts that received at least 100 replies each. This subset comprises 945,987 total replies, 19,152 of which were flagged by the original data providers, yielding an overall flagging rate of 2.02\%.

For each "flagged reply," we adopted the text classification labels provided by the original OAA source. The dataset also includes information on account verification status, follower counts, the number of flags on the original posts, and aggregated counts of flags across six distinct categories of replies. Accounts serve as the units of selection for the monitoring list, while replies serve as the units of counting for calculating "coverage of flagged content" and "conditional flagging rate." The same set of candidate accounts and the same definitions for "flagged" content are used throughout this study.

\subsection{Two Monitoring Objectives}

Let account \(i\) receive \(n_{i}\) replies, of which \(y_{i}\) are labeled. The candidate-account set is \(\mathcal{A}\) and the monitoring capacity is \(K\). The first method ranks accounts by \(y_{i}\) in descending order and selects \(K\) accounts. Its objective is to cover as many existing labeled replies as possible.

The second method selects accounts by conditional flagging rate. The raw rate \(y_{i}/n_{i}\) for a small account can be strongly affected by a small number of replies, so we rank accounts by the smoothed rate:

\[{\widehat{p}}_{i} = \frac{y_{i} + \tau p_{g}}{n_{i} + \tau},\quad\quad\tau = 50.\]

\(p_{g}\) is the average flagging rate of the comparison group. The full-sample ranking uses one overall mean; the group prior \(p_{g}\) in the stratified held-out experiment is calculated from ranking data in the corresponding group. Both methods use the same candidate set, and the main experiment sets \(K = 25\).

For the selected set \(S\) we calculate two measures:

\[\begin{aligned} Coverage(S)&= \frac{\sum_{i \in S}^{}y_{i}}{\sum_{i \in \mathcal{A}}^{}y_{i}},\\ Rate(S)&= \frac{\sum_{i \in S}^{}y_{i}}{\sum_{i \in S}^{}n_{i}}.\end{aligned}\]

"Coverage" is defined as the proportion of all flagged replies (across all candidate accounts) that are captured by the selected accounts S. The "conditional flagging rate" is defined as the proportion of flagged replies among all replies received by the selected accounts S. Both metrics are evaluated under the same constraint of K account slots. The smoothing parameter is fixed at 50. For full-sample rankings, the global flagging rate is used, whereas for stratified leave-out rankings, the prior mean for each group is calculated using only the data from that specific group.

\subsection{Account Grouping and Leave-Out Comparisons}

We categorize the 384 accounts into eight groups based on three criteria: verification status, whether the account has at least 100,000 followers, and whether at least one of the account's original posts has been flagged. The 25 available slots are allocated among these groups in proportion to the number of accounts in each, using the largest remainder method for rounding. Both the total-count-based ranking and the conditional-rate-based ranking employ identical group-specific quotas, ensuring that comparisons are made between sets of accounts with similar characteristics (Table 1).

\begin{table*}[t]
\centering
\footnotesize
\setlength{\tabcolsep}{3pt}
\begin{tabular}{@{}p{1.04in}p{1.02in}p{2.2in}p{2.22in}@{}}
\hline
\textbf{Verification status} & \textbf{Follower count} & \textbf{No labeled original posts: accounts / slots} & \textbf{Labeled original posts: accounts / slots} \\
\hline
Not verified & Below 100,000 & 30 / 2 & 240 / 16 \\
Not verified & At least 100,000 & 4 / 0 & 20 / 1 \\
Verified & Below 100,000 & 8 / 1 & 12 / 1 \\
Verified & At least 100,000 & 37 / 2 & 33 / 2 \\
Total & --- & 79 / 5 & 305 / 20 \\
\hline
\end{tabular}
\caption{Numbers of accounts and monitoring slots in each of the eight groups. Both methods use identical group-specific allocations, for a total of 384 accounts and 25 slots.}
\label{tab:1}
\end{table*}

For each account, historical counts are divided into flagged and unflagged components. A binomial random split---with a selection probability of 0.5---is applied to generate a ranking set (used for selection) and a held-out set (used for evaluation). Both methods rely exclusively on the ranking set to determine the final list, while the hold-out set is subsequently used to calculate coverage and conditional flagging rates. This process is repeated 500 times to derive the average hold-out results for a fixed set of accounts.

Additionally, we perform resampling with replacement of accounts within each of the eight groups. Each iteration involves re-splitting, selection, and evaluation (repeated 3,000 times) to assess how the effectiveness of the selected list varies based on the specific composition of the account sample. We also generate 3,000 datasets based on a null model---assuming homogeneous probabilities within each group---where the total number of replies and the group assignment for each account are preserved, while all accounts within the same group are assigned an identical probability of having a flagged reply. Both sets of data are subjected to the same slot allocation, ranking procedures, and hold-out evaluations, allowing for a direct comparison of results stemming from reply-count structure versus conditional characteristics.

\section{Empirical Results}

\subsection{High Content Coverage Corresponds to Different Account Selections}

Among the 384 candidate accounts, 25 were selected based on counts of flagged replies and another 25 based on smoothed conditional flagging rates; only one account appears on both lists. The list based on total counts includes 14,471 flagged replies, accounting for 75.56\% of all flagged replies, whereas the list based on conditional proportions includes 966 flagged replies, representing 5.04\%. The flagging rates within the two lists are 2.27\% and 7.13\%, respectively. Given the same number of slots, one list captures the majority of historically flagged responses, while the other concentrates on accounts with a higher proportion of flagged content.

Ranking by total count allocates slots to accounts that have already accumulated a large number of flags, thereby achieving a high cumulative coverage rate within a fixed number of slots. Ranking by conditional proportion, conversely, includes accounts that may have fewer total flags but a higher ratio of flags relative to their total volume of responses. With only one account overlapping between the two lists (when capped at 25 slots), the choice of metric for evaluating a list---such as content coverage---simultaneously determines which accounts are prioritized for monitoring. Table 2 illustrates how these selection outcomes shift as the number of slots varies (using 10 and 50 slots as examples).

\begin{figure*}[t]
\centering
\includegraphics[width=0.76\textwidth]{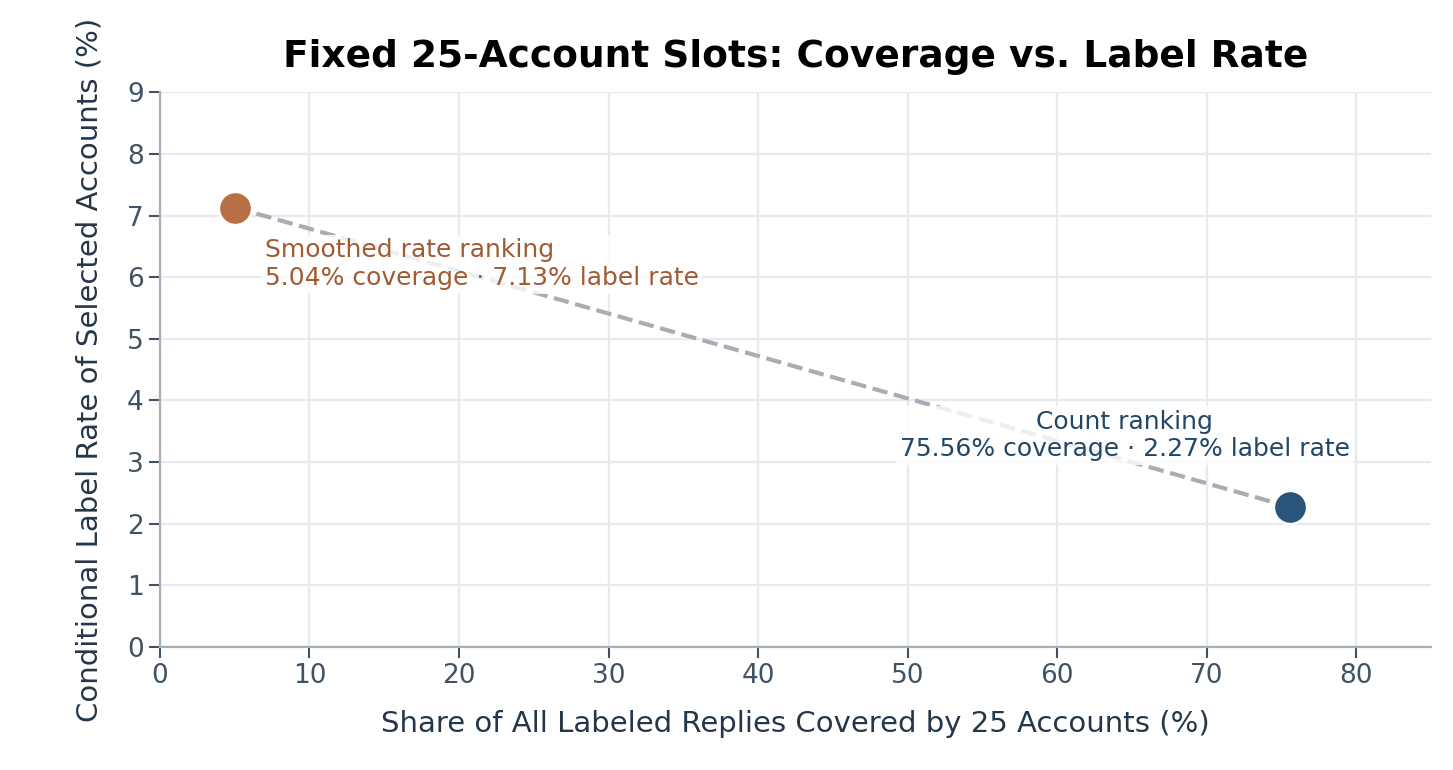}
\caption{For the same set of candidate accounts and a fixed number of slots, this chart shows the proportion of historical flagged replies covered by each ranking method (total count vs. conditional proportion), as well as the conditional flagging rates of the selected accounts.}
\label{fig:1}
\end{figure*}

\begin{table*}[t]
\centering
\footnotesize
\setlength{\tabcolsep}{3pt}
\begin{tabular}{@{}p{0.69in}p{0.55in}p{1.27in}p{1.27in}p{1.27in}p{1.27in}@{}}
\hline
\textbf{Monitoring slots} & \textbf{List overlap} & \textbf{Count-based coverage} & \textbf{Rate-based coverage} & \textbf{Count-based label rate} & \textbf{Rate-based label rate} \\
\hline
10 & 0 & 56.57\% & 2.18\% & 2.77\% & 9.28\% \\
25 & 1 & 75.56\% & 5.04\% & 2.27\% & 7.13\% \\
50 & 16 & 86.65\% & 26.51\% & 2.23\% & 4.90\% \\
\hline
\end{tabular}
\caption{Full-sample rankings under three account-slot capacities (K = 10, 25, 50), using a fixed smoothing strength of 50.}
\label{tab:2}
\end{table*}

\subsection{The "Long Tail" of Response Volume Can Lead to List Divergence}

The volume of responses is unevenly distributed across accounts. Even if accounts within each group share the same probability of a flagged reply, those with a higher volume of responses will naturally accumulate more flags; conversely, accounts with fewer responses may exhibit a higher proportion of flags simply due to random statistical fluctuation. Therefore, the low overlap between the two lists may, in itself, merely reflect the reply-count distribution and random variation.

We preserve the actual total reply count and group assignment for each account, while setting the flagging probability to be identical for all accounts within the same group. In 3,000 null model simulations, the two lists overlap by an average of only 0.50 accounts. Even when flagging probabilities are identical for all accounts within a group, selecting based on absolute totals versus conditional proportions can still result in lists that are almost entirely different.

However, the results from the held-out data in the homogeneous model tell a different story. The ratio of the held-out flagging rate for the "conditional proportion" list to that of the "total count" list averages 0.916, with a 95\% simulation quantile interval of 0.699--1.147. Although the two lists may be completely distinct, the conditional proportion list does not consistently exhibit a higher held-out flagging rate. In other words, the long-tailed reply counts can explain the separation of the lists; determining whether there are reproducible, conditional differences requires examining replies that were not included in the ranking process.

\subsection{Conditional flagging-rate differences persist in held-out historical replies}

In actual Online Abusive Attacks (OAA) data, both ranking methods utilize the same 384 accounts, the same eight-group quota structure, and the same held-out counts generated in each iteration. Across 500 random data splits, the average held-out flagging rate is 2.69\% for the "total count" list and 4.98\% for the "conditional proportion" list---the latter being 1.85 times higher than the former. The two lists share an average of 8.37 accounts. Even with fixed quotas per group, accounts selected via the conditional proportion method continue to show higher tagging rates among the replies that were not used for the ranking.

\begin{table*}[t]
\centering
\footnotesize
\setlength{\tabcolsep}{3pt}
\begin{tabular}{@{}p{3.5in}p{1.5in}p{1.65in}@{}}
\hline
\textbf{Held-out measure} & \textbf{Count-based list} & \textbf{Conditional-rate list} \\
\hline
Held-out label rate of selected accounts & 2.69\% & 4.98\% \\
Coverage of all held-out labeled replies & 48.67\% & 14.93\% \\
\hline
\end{tabular}
\caption{Average results from 500 random splits of historical counts. Both methods share the same account groupings, intra-group quotas, and held-out counts.}
\label{tab:3}
\end{table*}

The two lists also show different coverage results in the held-out part. The count-based list covers 48.67\% of all labeled replies in the held-out data, while the conditional-rate list covers 14.93\%. The conditional-rate list still has a higher proportion of labeled replies among its selected accounts, but these accounts cover fewer labeled replies in total than those on the count-based list. Comparing the two measures in the same held-out data shows the actual selection result corresponding to each monitoring slot.

Across 3,000 stratified account-level resamples, the average held-out label-rate ratio is 1.77, and the 95\% percentile interval is 1.38--2.18. The 95\% percentile interval for the difference in held-out label rates between the two lists is 1.26--2.88 percentage points. The mean ratio in the null model is 0.916, but the real data retain a positive difference under the same eight-group allocation. Reply-count distributions and recorded group attributes can affect list composition. But the real records also contain within-group conditional label-rate differences that can be observed again in held-out counts (Figure 2).

\begin{figure*}[t]
\centering
\includegraphics[width=0.76\textwidth]{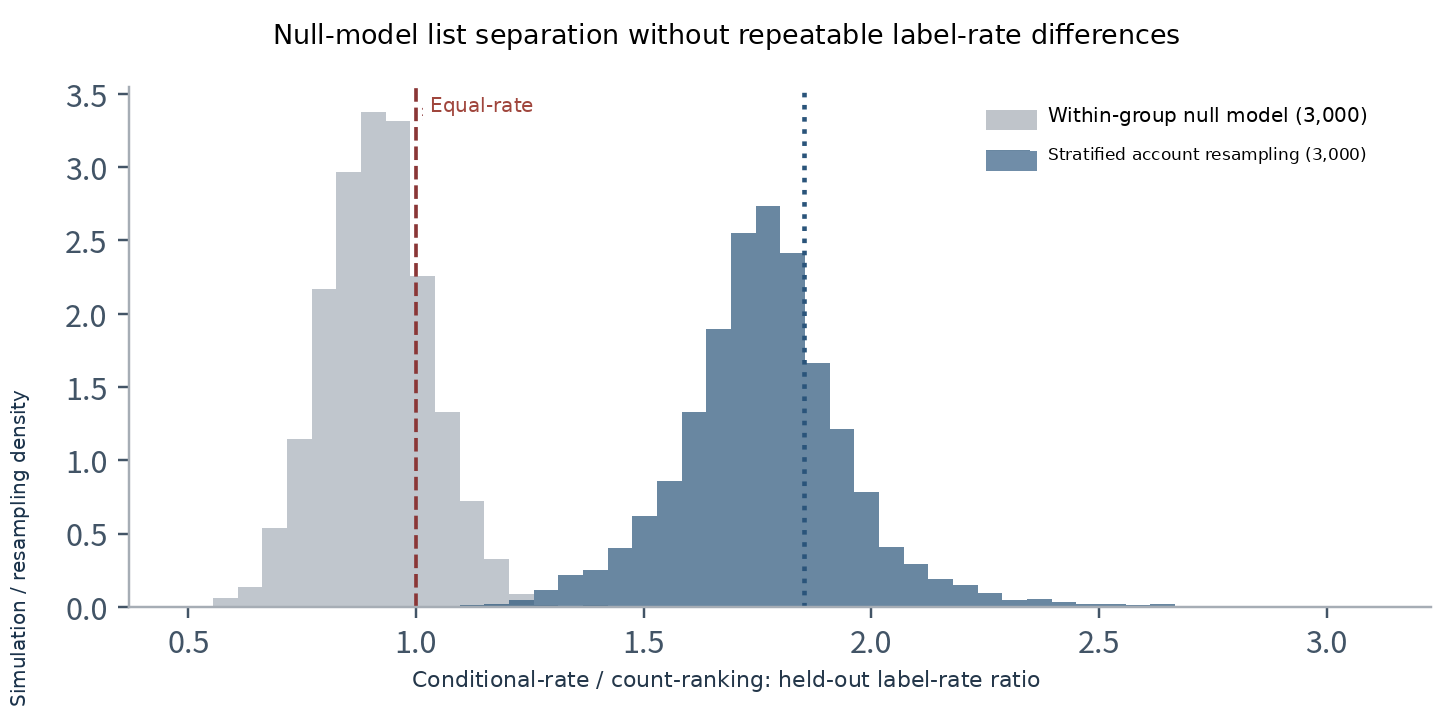}
\caption{Ratios of conditional-rate to count-based held-out label rates in 3,000 stratified account-level resamples of real data and 3,000 within-group homogeneous-probability null-model simulations. A ratio of 1 means the lists have equal held-out label rates.}
\label{fig:2}
\end{figure*}

\subsection{Selection rules also change label-category composition}

Among the six types of reply labels in OAA, the difference in the composition of the Identity\_Attack category is particularly clear. Of the 14,471 labeled replies on the count-based list, 281 belong to this category, or 1.94\%. Of the 966 labeled replies on the conditional-rate list, 174 belong to this category, or 18.01\%. The two lists use the same source classification but show different category compositions.

The count-based list covers more labeled replies in this category, but this category accounts for a smaller proportion of all labeled replies on that list. The corresponding number is smaller on the conditional-rate list, but its proportion is about 9.3 times as large. Therefore, which accounts receive monitoring slots changes not only the total number of labeled replies that can be recorded but also the proportion of each label category in the final monitoring sample.

\subsection{Sensitivity to monitoring capacity and smoothing strength}

We tested 12 combinations of monitoring capacity (K = 10, 25, 50) and smoothing strength ($\tau$ = 10, 25, 50, 100). Each configuration used the original eight-group quotas, 500 historical hold-out splits, 3,000 stratified account resamples, and 3,000 group-homogeneous null simulations. All configurations shared each random split. Fixed seeds distinguish these runs from the original experiment.

\begin{table*}[t]
\centering
\footnotesize
\setlength{\tabcolsep}{3pt}
\begin{tabular}{@{}p{0.55in}p{0.6in}p{1.42in}p{2.27in}p{1.35in}@{}}
\hline
\textbf{K} & \textbf{$\tau$} & \textbf{Held-out ratio} & \textbf{Bootstrap 95\% interval} & \textbf{Null-model mean} \\
\hline
10 & 10 & 1.79 & 1.40--2.33 & 0.93 \\
10 & 25 & 1.80 & 1.40--2.35 & 0.93 \\
10 & 50 & 1.81 & 1.41--2.37 & 0.93 \\
10 & 100 & 1.81 & 1.40--2.35 & 0.94 \\
25 & 10 & 1.86 & 1.37--2.23 & 0.92 \\
25 & 25 & 1.86 & 1.38--2.21 & 0.92 \\
25 & 50 & 1.85 & 1.37--2.21 & 0.92 \\
25 & 100 & 1.84 & 1.37--2.17 & 0.93 \\
50 & 10 & 1.74 & 1.39--2.27 & 0.96 \\
50 & 25 & 1.73 & 1.40--2.26 & 0.96 \\
50 & 50 & 1.72 & 1.39--2.25 & 0.96 \\
50 & 100 & 1.71 & 1.38--2.22 & 0.97 \\
\hline
\end{tabular}
\caption{Sensitivity of the ratio of held-out aggregate flagging rates (proportion-selected list / count-selected list). The bootstrap interval is the 2.5th--97.5th percentile range across 3,000 stratified account resamples. The null column is the mean ratio across 3,000 group-homogeneous simulations. The 500-split empirical ratio is the average of within-split ratios.}
\label{tab:4}
\end{table*}

The empirical held-out ratios range from 1.71 to 1.86 across the 12 configurations. Every bootstrap 95\% percentile interval has a lower endpoint above 1; the lowest is 1.37. Null-model mean ratios range from 0.917 to 0.965. The held-out advantage persists across the tested values of K and $\tau$.

\subsection{Independent UK election data: selection and forward tests}

We repeated the fixed-slot comparison on public daily aggregates of replies to UK election candidates {[}7{]}. Across 42 days, 2,581 accounts received 3,541,769 replies, including 157,844 labeled as abusive. Among 1,221 accounts receiving at least 100 replies, the count-based and smoothed-rate rankings shared eight of 25 monitored accounts (K = 25, $\tau$ = 50). The two lists covered 82.07\% and 43.50\% of labeled abuse among eligible accounts, respectively. In 500 random historical count splits, the average held-out rate ratio (rate list / count list) was 1.192 without group quotas and 1.085 with matched party-by-gender quotas; the latter ratio had a 2.5th--97.5th split-percentile range of 0.950--1.279. When rankings used only data before November 24, December 1, or December 8, the respective future rate ratios were 1.006, 0.969, and 1.213 without group quotas, versus 0.947, 0.970, and 0.897 with matched quotas.

\section{Discussion}

\subsection{Which accounts are missed under high content coverage}

The count-based list covers 75.56\% of the historical flagged replies with 25 accounts. This does not mean that all accounts with high conditional flagging rates enter the list. The count-based and proportion-based rules prioritize different monitoring outcomes.

Account slots and a budget for reviewing replies one by one are also different. When resources are allocated by account, each account occupies one slot, but the numbers of replies received by selected accounts may differ greatly. Therefore, when reporting content coverage, we also need to state which accounts with high label rates did not enter the list.

\subsection{Count structure and conditional differences behind list separation}

In the null model, accounts within the same group have the same labeling probability, but the two lists still overlap very little. This shows that differences in reply counts and random variation alone can cause the lists to separate. To determine whether proportion-based ranking finds a group of accounts with higher label rates, we also need to examine replies that were not used in ranking.

The held-out comparison using real data shows that the label rate of the proportion-based list remains higher than that of the count-based list. The same advantage does not appear in the null model. The two rankings use the same slots in the eight groups, reducing composition differences caused by verification status, follower count, and labels on original posts. The held-out data further show that differences in conditional label rates between real accounts can be observed again in another part of their historical replies.

\subsection{How account selection changes abuse-category statistics}

For the same type of abuse label, the number and proportion may differ between the two lists. The count-based list has 281 Identity\_Attack labels, or 1.94\% of all labeled replies on that list. The proportion-based list has 174 such labels, or 18.01\%. The proportion-based list has fewer labels in this category but a higher proportion, because the denominators of the two lists are different.

Therefore, 18.01\% describes the composition of labels on the proportion-based list, not the composition among all candidate accounts. The overall label rate of a list is also weighted by the number of replies. It is not equal to the simple average of the label rates of each selected account. To interpret these statistics, we need to state both the list-selection method and the denominator used in the calculation.

\subsection{Social impact and interpretation of historical data}

The consequences of online abuse may differ across accounts with unequal resources to respond. An account's low number of flagged replies does not by itself establish that the impact on its owner is small. Our dataset records reply labels, not financial losses or other individual harms.

The hold-out design randomly partitions historical counts, so it measures flagging-rate differences within those records. Evaluating next-period monitoring requires time-stamped replies. The 12 K $\times$ $\tau$ configurations test robustness within OAA; cross-platform generalization requires an independent dataset. Measuring economic or social harms requires corresponding outcome data.

The two rankings select different accounts in both datasets, but their held-out rate gap changes across settings. In the UK election data, the future rate ratio falls below one at all three cutoffs after matching party-by-gender quotas. Historical splits measure differences within the recorded period; chronological tests show how those differences carry into later replies.

\section{Conclusion}

Selecting accounts by the total number of flagged replies gives high coverage of previously observed abusive content. It can still exclude accounts whose replies contain a high proportion of flagged content. In the held-out historical records, proportion-based selection finds accounts with higher flagging rates, an advantage not reproduced by the within-group homogeneous null model. Monitoring evaluations should therefore report both flagged-reply coverage and which accounts are selected.

\section*{References}

{[}1{]} Alharthi, R., Alharthi, R., Shekhar, R., \& Zubiaga, A. (2023). Target-Oriented Investigation of Online Abusive Attacks: A Dataset and Analysis. IEEE Access, 11. \url{https://doi.org/10.1109/ACCESS.2023.3289148}

{[}2{]} Burke-Moore, L., Williams, A. R., \& Bright, J. (2025). Journalists are most likely to receive abuse: analysing online abuse of UK public figures across sport, politics, and journalism on Twitter. EPJ Data Science, 14, 41. \url{https://doi.org/10.1140/epjds/s13688-025-00556-8}

{[}3{]} Bandy, J., \& Lazovich, T. (2023). Exposure to Marginally Abusive Content on Twitter. Proceedings of the International AAAI Conference on Web and Social Media, 17(1), 24--33. \url{https://doi.org/10.1609/icwsm.v17i1.22123}

{[}4{]} Park, J. S., Seering, J., \& Bernstein, M. S. (2022). Measuring the Prevalence of Anti-Social Behavior in Online Communities. Proceedings of the ACM on Human-Computer Interaction, 6(CSCW2), Article 451. \url{https://doi.org/10.1145/3555552}

{[}5{]} Milne, D. N., McCabe, K. L., \& Calvo, R. A. (2019). Improving Moderator Responsiveness in Online Peer Support Through Automated Triage. Journal of Medical Internet Research, 21(4), e11410. \url{https://doi.org/10.2196/11410}

{[}6{]} Public repository and data files released by the original OAA authors: \url{https://github.com/RaneemAlharthi/Online-Abusive-Attacks-OAA-Dataset}

{[}7{]} Gorrell, G., Bakir, M. E., Roberts, I., Greenwood, M. A., \& Bontcheva, K. (2020). Which politicians receive abuse? Four factors illuminated in the UK general election 2019. EPJ Data Science, 9, 18. \url{https://doi.org/10.1140/epjds/s13688-020-00236-9}

\FloatBarrier

\section*{Generative AI Use Disclosure}
ChatGPT was used to help draft the second-dataset Results and Discussion additions and to prepare document formatting. The numerical results were checked against saved experiment outputs. The author is responsible for all scientific claims. 

\end{document}